\documentclass[aps,prx,reprint,runinaddress,superscriptaddress,amsmath,amssymb,floatfix,longbibliography]{revtex4-2}

\usepackage{amsmath,amssymb,bm,graphicx,color,gensymb,bbold,appendix}

\usepackage{microtype}
\usepackage{wasysym}
\usepackage{times}
\usepackage[varg]{txfonts}
\usepackage{mathrsfs}
\usepackage{bm}

\usepackage{xspace}
\usepackage{color}
\usepackage{graphicx}
\usepackage{booktabs}

\renewcommand{\vec}[1]{\boldsymbol{#1}}

\newcommand{\yzgo}{YbZnGaO$_4$\xspace}
\newcommand{\ymgo}{YbMgGaO$_4$\xspace}
\newcommand{\muB}{\mu_{\rm B}}

\definecolor{cred}{RGB}{188,55,84}

\begin{document}

\title{Phase Diagram of \yzgo in Applied Magnetic Field}

\author{William Steinhardt}
\affiliation{Department of Physics, Duke University, Durham, North Carolina, 27008, USA}
\author{P. A. Maksimov}
\affiliation{Bogolyubov Laboratory of Theoretical Physics, Joint Institute for Nuclear Research, Dubna, Moscow region, 141980, Russia}
\author{Sachith Dissanayake}
\affiliation{Department of Physics, Duke University, Durham, North Carolina, 27008, USA}
\author{Zhenzhong Shi}
\affiliation{Department of Physics, Duke University, Durham, North Carolina, 27008, USA}
\author{Nicholas P. Butch}
\affiliation{NIST Center for Neutron Research, National Institute for Standards and Technology, Gaithersburg, Maryland, 20899, USA}
\author{David Graf}
\affiliation{National High Magnetic Field Laboratory and Department of Physics, Florida State University, Tallahassee, Florida, 32310, USA}
\author{Andrey Podlesnyak}
\affiliation{Neutron Scattering Division, Oak Ridge National Laboratory, Oak Ridge, Tennessee, 37831, USA}
\author{Yaohua Liu}
\affiliation{Neutron Scattering Division, Oak Ridge National Laboratory, Oak Ridge, Tennessee, 37831, USA}
\author{Yang Zhao}
\affiliation{NIST Center for Neutron Research, National Institute for Standards and Technology, Gaithersburg, Maryland, 20899, USA}
\affiliation{Department of Materials Science and Engineering, University of Maryland, College Park, Maryland, 20742, USA}
\author{Guangyong Xu}
\affiliation{NIST Center for Neutron Research, National Institute for Standards and Technology, Gaithersburg, Maryland, 20899, USA}
\author{Jeffrey W. Lynn}
\affiliation{NIST Center for Neutron Research, National Institute for Standards and Technology, Gaithersburg, Maryland, 20899, USA}
\author{Casey Marjerrison}
\affiliation{Department of Physics, Duke University, Durham, North Carolina, 27008, USA}
\author{A. L. Chernyshev}
\affiliation{Department of Physics and Astronomy, University of California, Irvine, California 92697, USA}
\author{Sara Haravifard}
\affiliation{Department of Physics, Duke University, Durham, North Carolina, 27008, USA}
\affiliation{Department of Mechanical Engineering and Materials Science, Duke University, Durham, North Carolina 27708, USA}
\date{\today}

\begin{abstract}
Recently, Yb-based triangular lattice antiferromagnets have garnered significant interest as possible quantum spin liquid candidates.  One example is \ymgo, which showed many promising spin liquid features, but also possesses a high degree of disorder owing to site-mixing between the non-magnetic cations.  To further elucidate the role of chemical disorder and to explore the phase diagram of these materials in applied field, we present neutron scattering and sensitive magnetometry measurements of the closely related compound, \yzgo. Our results suggest a difference in magnetic anisotropy between the two compounds, and we use key observations of the magnetic phase crossover to motivate an exploration of the field- and exchange parameter-dependent phase diagram, providing an expanded view of the available magnetic states in applied field. This enriched map of the phase space serves as a basis to restrict the values of parameters describing the magnetic Hamiltonian with broad application to recently discovered related materials.
\end{abstract}

\maketitle

\section{Introduction}
Recent years have seen renewed interest in triangular-lattice antiferromagnets featuring anisotropic interactions and other traits conducive to exotic quantum ground states, particularly in the hunt for experimental realizations of quantum spin liquids.  Mapping the phase diagrams of these materials is thus of paramount importance, as the variation in magnetic anisotropy, relative  exchange coupling strengths, and corresponding magnetic Hamiltonians offered by different materials are fundamentally important to the pursuit of such states.  

One system that sparked an explosion of experimental and theoretical interest is the triangular-lattice antiferromagnet \ymgo, which has led to a broad research effort that currently involves a number of related compounds  \cite{sanders2017magnetism,baenitz2018planar,sichelschmidt2020effective,sichelschmidt2019electron,cevallos2018anisotropic,li2018absence,liu2018rare,ding2019gapless,ranjith2019field,ranjith2019anisotropic,xing2019field,bordelon2019field,xing2019synthesis,xing2019class,scheie2020crystal,bastien2020long}.  There are two important physical aspects of this group of materials that have become clear due to recent insights. Because of the localized nature of the $f$-orbitals, the ranges of the effective spin-spin interactions in these insulating materials are strongly limited, implying that all of the Kramers-ion-based materials are expected to be closely described by the same nearest-neighbor anisotropic-exchange model with the parameters that are permitted by triangular-lattice symmetry  \cite{li2016anisotropic,rau2018frustration,zhu2018topography,maksimov2019anisotropic}. This reasonably compact model should provide a consistent interpretation of the current and future experiments and give important new insights into fundamental properties of these materials \cite{li2020spin}.   

The second generic feature is the strong effect of disorder observed in some of the well-studied representatives of these compounds. It has been argued theoretically that even a benign form of bond disorder necessarily generates perturbations that are relevant in the Imry-Ma sense  \cite{imry1975random,zhu2017disorder,parker2018finite}, making a consideration of the defects an inevitable and essential part of a realistic description of most anisotropic-exchange magnets.  Empirically, many of the newly synthesized materials seem to show no magnetic ordering  \cite{sanders2017magnetism,baenitz2018planar,sichelschmidt2020effective,sichelschmidt2019electron,cevallos2018anisotropic,li2018absence,liu2018rare,ding2019gapless,ranjith2019field,ranjith2019anisotropic,xing2019field,bordelon2019field,xing2019synthesis,xing2019class,scheie2020crystal,bastien2020long}. Then, given the strong disorder effects, it is a question whether the non-magnetic ground states of these materials are due to a genuine quantum-disordered spin-liquid (SL) phase, or due to a scenario similar to the disorder-induced “spin-liquid mimicry,” suggested for \ymgo  \cite{zhu2017disorder}.  

There is also an intriguing broader question on whether the disorder-induced spin-liquid-like behavior retains any of the unique and desired properties of the intrinsic spin liquids, thus potentially turning disorder into a feature rather than an obstacle  \cite{kimchi2018valence,andrade2018cluster,bilitewski2017jammed}. A counterintuitive example of the role of disorder in a related material is the case of NaYbO$_2$, where introducing Na$^+$ site vacancies leads to an antiferromagnetic transition at a few Kelvin and thus disorder supports an ordered phase\cite{guo2020magnetic}. Further understanding and insight into the role of disorder are needed to make progress in this direction.  

One of the persistent issues in the studies of the anisotropic-exchange magnets in general and of the rare-earth family in particular is the identification of their model parameters  \cite{maksimov2020rethinking}. In the case of the disorder-induced pseudo-SL state, this problem is further aggravated, as it is not clear what  state the disorder-free system would have assumed. In this work, we propose that the experimental and theoretical investigations of the \emph{field-induced phases} offers a powerful instrument to significantly narrow the allowed  parameter space to a region that is consistent with the material's phenomenology. 

In \ymgo  and \yzgo, the source of disorder is due to the $R\bar{3}m$ symmetry, leading to fifty-fifty site mixing of the non-magnetic cations Mg$^{2+}$ or Zn$^{2+}$ with the Ga$^{3+}$.  Efforts to determine the exchange parameters for placing \ymgo in proposed phase diagrams and otherwise comparing to the numerous theoretical investigations to affirm or deny a QSL state were obstructed by the various broadening effects and consequentially enhanced uncertainty in the measurements. Several studies concentrated their efforts on further refining measurements of the exchange parameters \cite{zhang2018hierarchy,steinhardt2020fieldinduced}.

In this work, we provide a detailed study of the field-induced effects and characteristics of YbZnGaO$_4$  and offer comparison with the results in YbMgGaO$_4$  \cite{steinhardt2020fieldinduced}.  Informed by measurements from high resolution magnetometry and a variety of neutron scattering techniques, we put forward a theoretical analysis of the structure of their field-induced phase diagram and propose a parameter region for both materials that is compatible with our empirical findings.  

\section{Results}
\begin{figure*}[t]
\includegraphics[width=\linewidth]{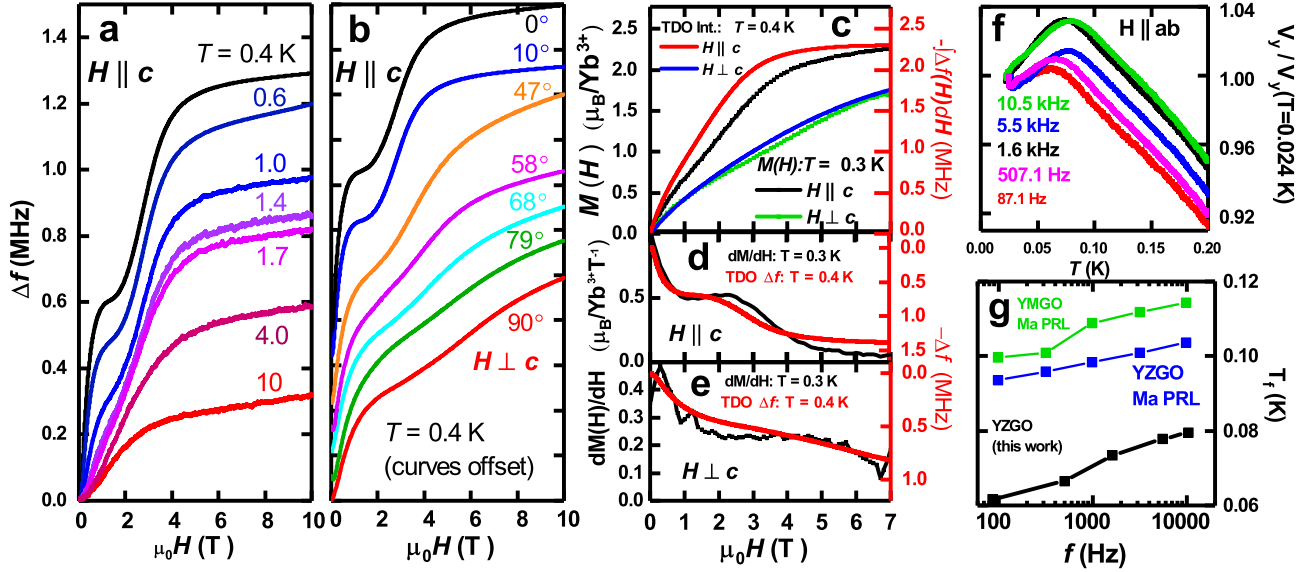}
\vskip -0.2cm
\caption{Tunnel diode oscillation (TDO) frequency and ac susceptibility. (a) and (b) TDO ($\Delta f/f \propto \Delta \textbf{M}/ \Delta \textbf{H}$) shows an anomaly for $\textbf{H}\parallel c$ and  $\textbf{H}\perp\textbf{c}$ respectively which weakens as temperature increases (curves offset for clarity in (b)). (c) Anomaly’s response to applied field shows anisotropy.  (e) $d\textbf{M}/d\textbf{H}$ measured with SQUID corroborates TDO measurement. (f) integrating TDO $\Delta$frequency from 0 to the approximate saturation is further corroborated by dc susceptibility measurements. (g) Comparison of critical temperatures measured in this work for YbZnGaO$_4$ to earlier measurements by Ma et al. \cite{ma2018spin}} 
\label{TDO_SQUID}
\end{figure*}

\subsection{Preliminary characterization}
Susceptibility measurements on a 1.81 mg single crystal sample of YbZnGaO$_4$ conducted using an in-house Cryogenic S700X SQUID magnetometer (with 3He probe) yield low-temperature fits suggesting Curie-Weiss temperatures of $\Theta_{CW}=-2.67$ K and $\Theta_{CW}=-2.62$ K for parallel and perpendicular to the sample $\vec{c}$ axis, respectively (see Supplementary Figure 4).  These values are more isotropic than those reported for \ymgo or for \yzgo in earlier works though \yzgo was also more isotropic than \ymgo in those measurements)  \cite{ma2018spin}.

We emphasize that while the difference in the Curie-Weiss temperature for the in-plane vs out-of-plane field has initially suggested a rather strong easy-plane character of \ymgo  \cite{paddison2017continuous}, the subsequent spectroscopic studies have hinted at a rather moderate $XXZ$ anisotropy, yielding a nearly Heisenberg value of $\Delta\!=\!0.8$--0.9  \cite{zhang2018hierarchy}.  In the case of \yzgo,  Curie-Weiss temperatures for the in-plane and out-of-plane fields from our measurements and those of earlier works  \cite{ma2018spin}  are  much closer to the Heisenberg limit. 
Given the trend, this indicates that the anisotropy  in \yzgo may, in fact, be of the easy-axis type. A more direct demonstration of that is offered by the results that are discussed in Sec.~II.B. 

A unique opportunity afforded by a close comparison between \ymgo and \yzgo is the qualitative contrast of the effect of the cation substitution on the disorder between the two materials. 
One obvious consideration is the differing ionic radii of the Mg$^{2+}$ and Zn$^{2+}$, which are approximately 72 and 74 pm respectively (a 2.7$\%$ difference).  This small difference yields slightly smaller lattice parameters for \ymgo when compared to \yzgo, as shown by comparison of parameters given by references  \cite{li2015gapless} and  \cite{ma2018spin}, respectively, which, in turn, may yield marginally stronger exchange for \ymgo, also explaining the smaller Curie-Weiss temperatures and lower field onset for the anomaly discussed in Sec.~II.B for \yzgo.  This is consistent with the observation that the related compound NaYbO$_2$, with an in-plane lattice parameter smaller by only about 1.8$\%$ from YbMgGaO$_4$, shows a significantly larger Curie-Weiss temperature ($\Theta_{CW} = -10.38$ K\cite{bordelon2019field} as opposed to ~$-4$ K for \ymgo)\cite{steinhardt2020fieldinduced}.  Furthermore, both Zn$^{2+}$ and Ga$^{3+}$ have $d^{10}$ electronic configurations, whereas Mg$^{2+}$ is $p^6$.  While the displacement of Yb$^{3+}$ can still be expected based on the charge difference between the cations, leading to the observed broadening in inelastic neutron scattering studies of the single magnon dispersion and crystal electric field levels \cite{li2017crystalline}, the local environment may be more homogeneous for \yzgo, and may be related to the difference in anisotropic response under field. However, further studies will be required to compare the effective role of disorder between the two systems.

\subsection{Tunnel diode oscillator technique and SQUID magnetization}

The first indications of the magnetic transitions in \yzgo, which are similar to the ones we have previously identified in \ymgo \cite{steinhardt2020fieldinduced}, were found using the tunnel diode oscillator (TDO) technique (see methods).  As the applied magnetic field is changed, the sample magnetization $M (\vec{H})$ with respect to field is altered, thus changing the inductance of the coil and the measured resonant frequency of the circuit, yielding a signal proportional to $\chi (\vec{H})$.  From Figure \ref{TDO_SQUID} (a), where the change in resonant frequency is plotted versus the applied field, a clear nonlinearity is apparent beginning just below 1 T.  This nonlinearity persists to at least 2 T for the field parallel to sample $\vec{c}$ axis.  Upon raising the temperature, the feature is completely suppressed at about 4 K, affirming its magnetic origin. This behavior is also consistent with a  similar feature in \ymgo \cite{steinhardt2020fieldinduced}.  As the sample orientation is rotated with respect to the field, the anisotropic response is made apparent, with the feature encompassing a broader range of the field for $\textbf{H}\perp\textbf{c}$. This feature is confirmed by more conventional magnetization and susceptibility measurements carried out using the in-house SQUID magnetometer with a 3He probe in temperatures down to 300 mK. The same distinct plateau-like feature is apparent in the $\chi(\vec{H})$ in both TDO and SQUID measurements, as is clear from Figures \ref{TDO_SQUID} (d) and (e).  Integrating the change in frequency with respect to applied field yields a curve consistent with magnetization as measured in SQUID (see Figure \ref{TDO_SQUID} (c)).  We note that the anomaly in \yzgo measured via  TDO and SQUID magnetization occurs at a slightly lower field  compared to the analogous feature measured in \ymgo \cite{steinhardt2020fieldinduced} (see Supplementary Figure 5).

The features in the magnetization derivative and tunnel diode oscillator technique (TDO) 
measurements in both in-plane and out-of-plane fields  are reminiscent of the plateau-like behavior that 
is expected in the canonical Heisenberg or $XXZ$ nearest-neighbor triangular-lattice magnets \cite{starykh2015unusual}.
Importantly, the TDO and magnetization measurements in \yzgo and \ymgo, together with the 
Curie-Weiss extrapolations from the susceptibility mentioned above, suggest an important distinction between the two materials.
The effects of the in-plane and the out-of-plane field directions show different behavior as a function of the field orientation - compare Figure~\ref{TDO_SQUID}(b) of this work and Figure~1(c) in Ref.~\citep{steinhardt2020fieldinduced}. Specifically, the  relative severity of the anomaly in TDO data in \yzgo in $H\!\parallel\! c$ are resemblant of that in $H\!\perp\! c$ data of \ymgo, and vice versa. This suggests that \yzgo and \ymgo correspond to different types of the $XXZ$ anisotropy, 
the easy-axis and the easy-plane, respectively.

\subsection{ac susceptibility and disorder}

From our measurements of ac susceptibility of \yzgo (see Figure \ref{TDO_SQUID} (f)), we find critical temperatures corresponding to the characteristic cusps of the ac susceptibility occur at substantially lower temperatures than previously measured  \cite{ma2018spin}.  Indeed, our characteristic temperatures are around $20\%-30\%$ lower for all comparable frequencies - see Figure \ref{TDO_SQUID} (g).  Consequently, for our measurement of $\Delta P = \frac{\Delta T_f}{T_f \Delta \log(\omega)}$, a quantitative measure of the freezing temperatures per decade of characteristic temperatures, we find a value of $\Delta P = 0.139(4)$, substantially larger than the $\Delta P = 0.053(4)$ of previous work.  This value of $\Delta P$ is typically associated with superparamagnetic behavior  \cite{mydosh1993spin} as opposed to spin glasses.  That being said, insulating spin glasses typically show greater frequency dependence~\cite{mydosh1993spin}.  If this $\Delta P$ \emph{is} interpreted as indicative of superparamagentic behavior, it may be understood as a consequence of many microscopic domains with insignificant cooperative freezing.  This phenomenon is likely due to  disorder, but further suppression of the freezing temperature could be attributed to the high degree of frustration as well as disorder.  The percentage of frozen spins is estimated to be approximately 16\% (see Supplementary Figure 9), comparable to the previous study \cite{ma2018spin} and to the case of \ymgo \cite{paddison2017continuous}. 

General questions about the effects of disorder in spin systems persist, especially in light of its aforementioned potential relationship to QSL states for some materials \cite{kimchi2018valence,andrade2018cluster,bilitewski2017jammed}.  The origin and effects of disorder related to the observed spin-liquid features of \ymgo have been addressed earlier \cite{zhu2017disorder,li2017crystalline}.

\subsection{Neutron scattering}

\begin{figure*}

\includegraphics[width=\linewidth]{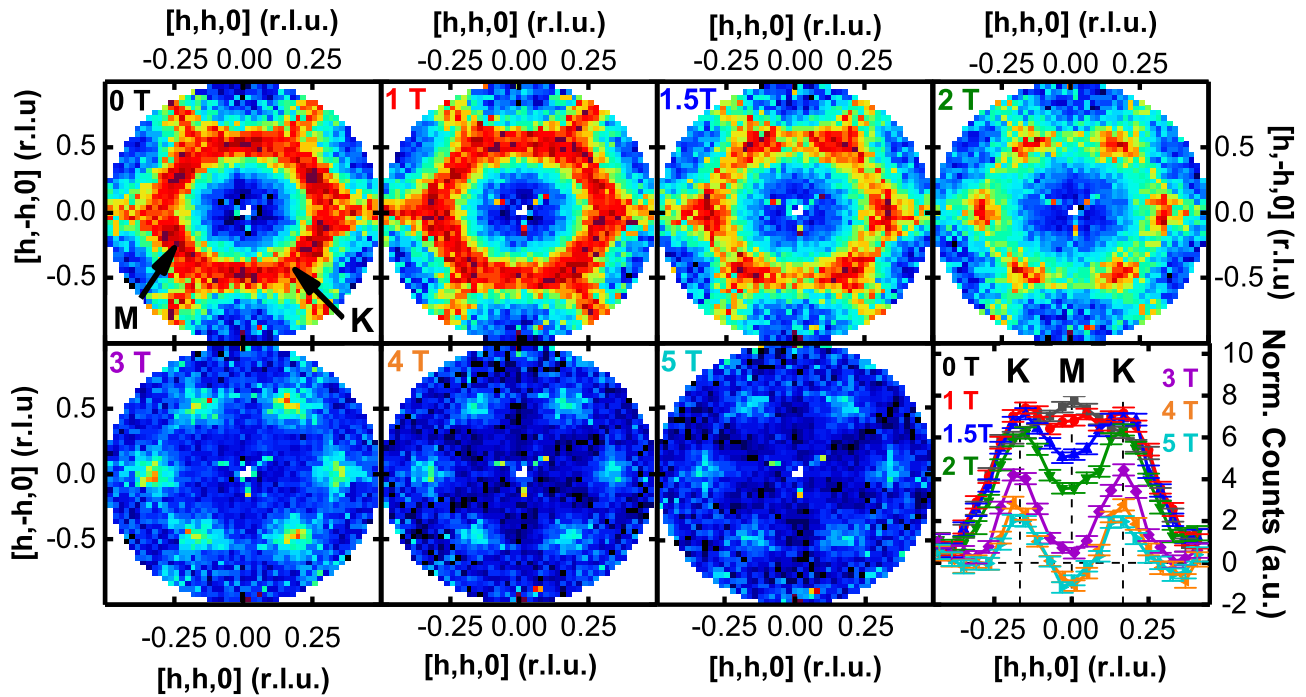}
\vskip -0.2cm
\caption{Diffuse neutron scattering. First seven panels show evolution of diffuse neutron scattering with increasing applied field for $\vec{H} \parallel \vec{c}$ for data integrated over $0.5 < L < 0.5$.  The eighth panel shows integrations of line cuts across the first BZ edge, where the uncertainty represents one standard deviation.  Sample temperature was 130 mK and a 20 K background was subtracted to isolate magnetic contributions.} 
\label{CORELLI}
\end{figure*}

 For this work, diffuse neutron scattering data were collected at CORELLI at Oak Ridge National Laboratory in total scattering mode (see Figure \ref{CORELLI}).  The sample was aligned with the $[h,k,0]$ scattering plane, and applied field along the sample $\vec{c}$ axis.  Data were collected at 130 mK for 0, 1, 1.5, 2, 3, 4, and 5 T.  Color maps of the neutron scattering intensity after subtracting the 20 K as background reveal the evolution of the magnetic structure, which is qualitatively comparable to the \ymgo data.  With no applied field, the intensity largely resides at the high-symmetry M points on the edges of the Brillouin zone (see also Supplementary Figure 6, left panel).  At 1 T the scattering intensity is almost completely uniform along the zone edge, while at 1.5 T the intensity is found predominantly at the high-symmetry K points.  Intensity at the M points is well established to correspond to stripe-ordered states in long-range ordered triangular systems, while intensity at the K point is generally suggestive of 120$^\circ$-type ordering or other three-sublattice states \cite{maksimov2019anisotropic}.  As in the case of \ymgo, this migration of the scattering intensity with applied field corresponds to the anomaly observed in the magnetometry data.  The changes in intensity were further confirmed by measurements with the triple-axis spectrometer at SPINS (see Supplementary Figure 7).

We further measured inelastic neutron scattering (INS) from a \yzgo single crystal sample at CNCS\cite{ehlers2011cncs} in applied field(see Figure \ref{CNCS}) at Oak Ridge National Laboratory.  Here we again see a clear evolution of the intensity as a function of energy and $Q$ with increasing applied field. We note that at low field the intensity is notably concentrated on the M points.  The intensity has no clear dispersion up to about 3 T, but instead consists of a broad continuum, similar to the 0 T with weakening intensity as the spins are presumably canted further from the scattering plane with increased field. At 4 T, the scattering remains broad in energy, but a dispersion is faintly visible at low energy along the zone edge.  This dispersion has minima at the zone edges and rises as it approaches the $\Gamma$ points.  The shape of this dispersion closely resembles what is measured in the polarized state measured at 8 T, indicating that the system is approaching polarization. Measurements at DCS\cite{COPLEY2003477} (see Supplementary Figure 6) and SPINS (see Supplementary Figure 7) at the National Institute of Standards and Technology confirm the features shown in Figure \ref{CNCS} in a diversity of backgrounds and instrument setups. At 8 T the system is nearly completely polarized, and a clear dispersion is evident.  As in the case of \ymgo \cite{paddison2017continuous}, the dispersion is broadened, likely due to the disorder and the resulting distribution of exchange parameters and $g$-factors.  

Additional measurements to characterize \yzgo's response to applied field were conducted using polarized neutrons at BT7\cite{lynn2012bt7} at the National Institute of Standards and Technology, with a vertical guide field set up (see Supplementary Figure 8).  After subtracting background measurements (40 K) from base (0.3 K) and correcting for the polarization rate, comparison of the spin flip (SF, measuring the in-plane component) shows a stronger in-plane component along the zone edge at 0 T, compared to 2 T, with particularly high intensity in the vicinity of the high-symmetry M point, likely affirming increased canting of the spins with increasing field, but also possibly indicating a reduced spin component parallel to the zone edge (pointing to nearest neighbors in real space).  This greater intensity near the zone edge for the SF scattering at 0 T can also be seen in the orthogonal cut from M to $\Gamma$. 

\subsection{Model and zero-field phases}
The interplay of the crystal field and spin-orbit coupling on the magnetic moment of the Kramers ion results in the splitting of its levels into a well-separated doublet structure 
built from a mix of various spin and orbital states. The exchange interactions of the lowest doublets (pseudo-spins-$\frac{1}{2}$) are constrained only by the discrete lattice symmetry. For the triangular lattice of \ymgo and \yzgo, this general anisotropic-exchange nearest-neighbor model  is given by
\begin{align}
\label{HJpm}
{\cal H}=&\sum_{\langle ij\rangle}\Big\{J \Big(S^{x}_i S^{x}_j+S^{y}_i S^{y}_j+\Delta S^{z}_i S^{z}_j\Big)\\
+&2 J_{\pm \pm} \Big[ \Big( S^x_i S^x_j - S^y_i S^y_j \Big) \cos\tilde{\varphi}_\alpha-
\Big( S^x_i S^y_j+S^y_i S^x_j\Big)\sin\tilde{\varphi}_\alpha \Big]\nonumber\\
 +&J_{z\pm}\Big[ \Big( S^y_i S^z_j +S^z_i S^y_j \Big) \cos\tilde{\varphi}_\alpha 
 -\Big( S^x_i S^z_j+S^z_i S^x_j\Big)\sin\tilde{\varphi}_\alpha \Big]\Big\},\nonumber
 \end{align}
where bond angles $\tilde{\varphi}_\alpha$ are that of the primitive vectors of the lattice with the $x$ axis, $\tilde{\varphi}_\alpha\!=\!\{0,2\pi/3,-2\pi/3\}$. The $J_{\pm \pm}$ and $J_{z\pm}$ bond-dependent terms are due to the strong spin orbit coupling\cite{li2016anisotropic}.

The zero-field phase diagram of the model \eqref{HJpm} with an additional second-nearest-neighbor exchange $J_2$ has been studied extensively  \cite{li2016anisotropic,zhu2018topography,maksimov2019anisotropic,liu2016semiclassical,luo2017ground,rousochatzakis2016kitaev,iaconis2018spin}, and its 3D classical version is shown in Figure~\ref{fig_3d} in the  $J_2$-$J_{\pm\pm}$-$J_{z\pm}$ axes for $\Delta=1$, with all couplings in units of  $J>0$.  

There are three main ordered states in the antiferromagnetic limit of the $XXZ$ interaction: a coplanar three-sublattice $120^\circ$ state, which corresponds to the ordering vector at $K$ points, a collinear stripe-\textbf{x}  two-sublattice state with spins pointing along the nearest-neighbor bonds and the ordering vector at $M$ points, and a second collinear stripe-\textbf{yz} state with the same ordering vector but spins tilted out of the lattice plane and perpendicular to the nearest-neighbor bonds.

We should note that the phase diagram in Figure~\ref{fig_3d} is simplified. The simplification comes from only taking the single-\textbf{Q} spiral ansatz that does not include more complicated multi-\textbf{Q} states  \cite{luo2017ground,iaconis2018spin}. Moreover, the quantum version of the phase diagram also has a spin-liquid phase  \cite{zhu2018topography,maksimov2019anisotropic}, which is located along the tricritical boundary between stripe and $120^\circ$ states for a limited range of the $XXZ$ anisotropy near the Heisenberg limit. 

\subsection{Exploring the XXZ parameter space}
One of the puzzling features observed in some of the first experiments in \ymgo  \cite{li2015gapless} was an indication of the field-induced phase crossovers, seen in magnetic susceptibility. Recent susceptibility and TDO measurements of \ymgo have further supported these observations   \cite{steinhardt2020fieldinduced}. Here we present a variety of measurements on high quality single crystal samples to show that very similar features, indicating field-induced crossovers for both $H\!\parallel\! c$ and $H\! \perp\! c$ also occur in \yzgo.  

The neutron scattering measurements in the out-of-plane magnetic field $H\!\parallel\! c$  \cite{steinhardt2020fieldinduced}, also indicated a field-induced crossover and brought another piece of evidence to light. Neutron diffraction has showed that the field-induced crossover is accompanied by a shift of magnetic intensity from the $M$-points at the lower fields to $K$-points at the higher fields. In an ordered state, such an intensity shift would correspond to a transition from a four-sublattice to a three-sublattice state.  Our key finding is that this feature alone allows one to put strong boundaries on the exchange parameters of the system when the phase diagram of relevant parameters is considered. 

From the susceptibility data presented above and as established by earlier work in the case of \ymgo, \yzgo can be characterized as easy-axis anisotropy while \ymgo is easy-plane. Therefore, in the following we use two representative values of $XXZ$ anisotropy for the model (\ref{HJpm}), the easy-plane case $\Delta\!=\!0.8$ that is  related to \ymgo, and the easy-axis case $\Delta\!=\!1.1$ as related to \yzgo.

First, we explore the parameter region that would allow for a transition from the four-sublattice to the three-sublattice state at some finite value of the  out-of-plane magnetic field $H_c$ using classical energy minimization.  We fix the second-nearest-neighbor coupling $J_2$ to the value $0.05J$ (red line in Figure~\ref{fig_3d}) and the $XXZ$ anisotropy in the model (\ref{HJpm}) to the two values discussed above. That leaves the bond-dependent anisotropies $J_{\pm\pm}$ and $J_{z\pm}$ and the field as the parameters to scan through. We highlight our findings in Figure~\ref{HcHsJzpJp} (a) and (b) in the form of an intensity plot of the field $H_c$ of such a four-to-three sublattice transition in units of the saturation field, $H_c/H_s$, and in the $J_{\pm\pm}$--$J_{z\pm}$ axes (in units of $J$). The $120^\circ$ phase is a three-sublattice state already at zero field and remains such  for all the fields, while most of the stripe phase regions remain four-sublattice throughout the entire field range. 
 
Our key finding is illustrated by the gradient-color regions interpolating between the ``only-three-'' and ``only-four-sublattice"  regions in Figure~\ref{HcHsJzpJp} (a) and (b). They demonstrate that already at the level of our classical energy analysis, the four-to-three sublattice transition indeed takes place at some value of $H_c\!<\!H_s$ in a surprisingly extended region that emanates from the   $120^\circ$ part of the phase diagram into the stripe-\textbf{yz} phase and extends up to $J_{z\pm} \!\sim\! J$ in both cases, with the intensity emphasizing how far this transition is from the saturation field $H_s$.  

It should be noted that for $S=1/2$, quantum effects in zero field are known to broaden the stability region of the $120^\circ$ phase beyond the boundaries of the classical model consideration  \cite{zhu2018topography}. Therefore, one may also expect  the region of the four-to-three sublattice transition to extend beyond the classical predictions in this work.

While one can expect that quantum effects will further stabilize and extend  the field-induced three-sublattice states, we note that experimentally the ``4-to-3''  (or M-to-K) transition in both \ymgo and \yzgo  occurs at a rather low field, $H_c \!< \!0.5 H_s$, which provides further restrictions on the possible parameter ranges. At the level of our approximations, this constraint  puts \ymgo and \yzgo in a close proximity to the $120^\circ$ phase boundary, giving the upper bound $J_{z\pm}/J \!\alt\! 0.4$. 

As one can see in Figure~\ref{HcHsJzpJp} (a), there is a narrow region inside the stripe-\textbf{x} phase where the transition of interest also occurs, but it involves another transition back to the four-sublattice state at higher fields, so we do not consider it as a relevant region to the materials in question here.

\begin{figure*}
\includegraphics[width=\linewidth]{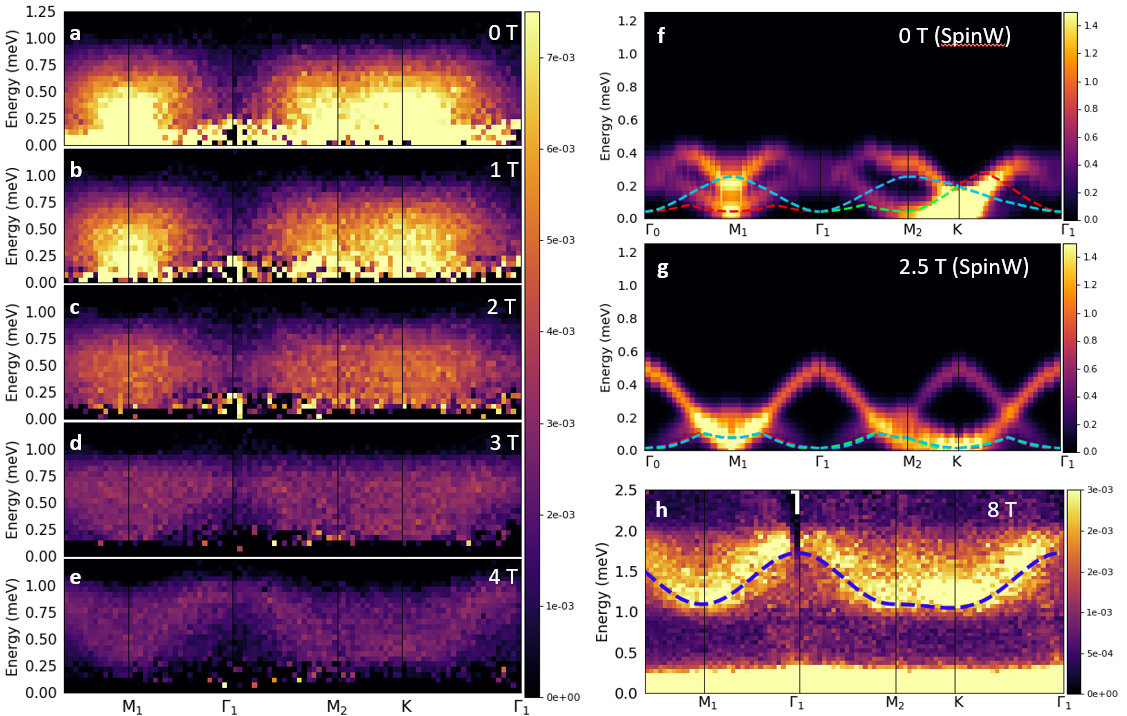}
\vskip -0.2cm
\caption{Inelastic neutron scattering. Panels (a-e) show evolution of inelastic neutron scattering with increasing applied field for $\vec{H} \parallel \vec{c}$ for integer fields from 0 to 4 T, where the 8 T data (where magnetic excitations have been lifted to above 1.2 meV) has been used for background subtraction).  Panels (f) and (g) are calculated using SpinW for 0 T (stripe yz) and 2.5 T (V-state), respectively, with a summation from the possible magnetic domain orientations to best compare to the short-range order observed in experiment. Parameters used for SpinW calculations are (in meV) $J=0.15$, $J_{\pm\pm} = -0.0045$, $J_{z\pm} = 0.045$, $J_{zz}=0.165$, $J^2=0.0075$, $J_{zz}^2=0.00825$, $g_{\parallel}=3.82$, such that  $J_{\pm\pm}/J = -0.03$, and $J_{z\pm}/J=0.3$. The dashed lines of (f) and (g) indicate the minimum energy for the 2-magnon dispersion for the possible domains. (h) shows 8 T dispersion, and a dotted line indicating the calculated curve from LSWT for the parameters described above. Experimental data are shown for an integration perpendicular to the path direction of width Q = 0.36 \r{A}$^{-1}$. Cuts for experimental data were generated using Horace\cite{Ewings2016132}. }
\label{CNCS}
\end{figure*}

\begin{figure}[t]
\includegraphics[width=0.7\linewidth]{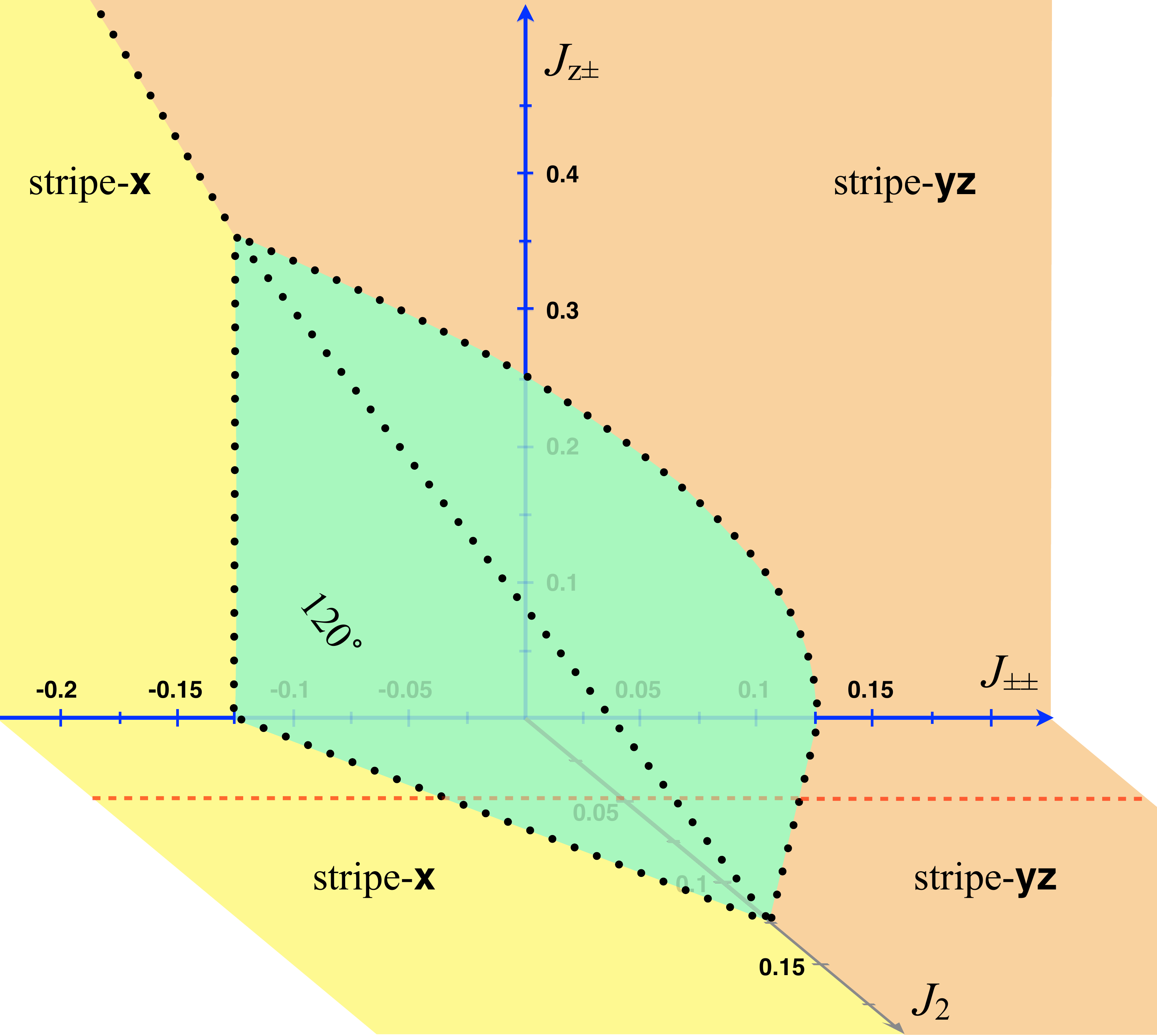}
\caption{A simplified classical zero-field $J_2$-$J_{\pm\pm}$-$J_{z\pm}$ phase diagram of the model (\ref{HJpm}). 
Dashed red line marks a representative value of the $J_2$ term used for all panels of Figure~\ref{HcHsJzpJp}}.
\label{fig_3d}
\end{figure}
\begin{figure*}[b]
\includegraphics[width=0.75\linewidth]{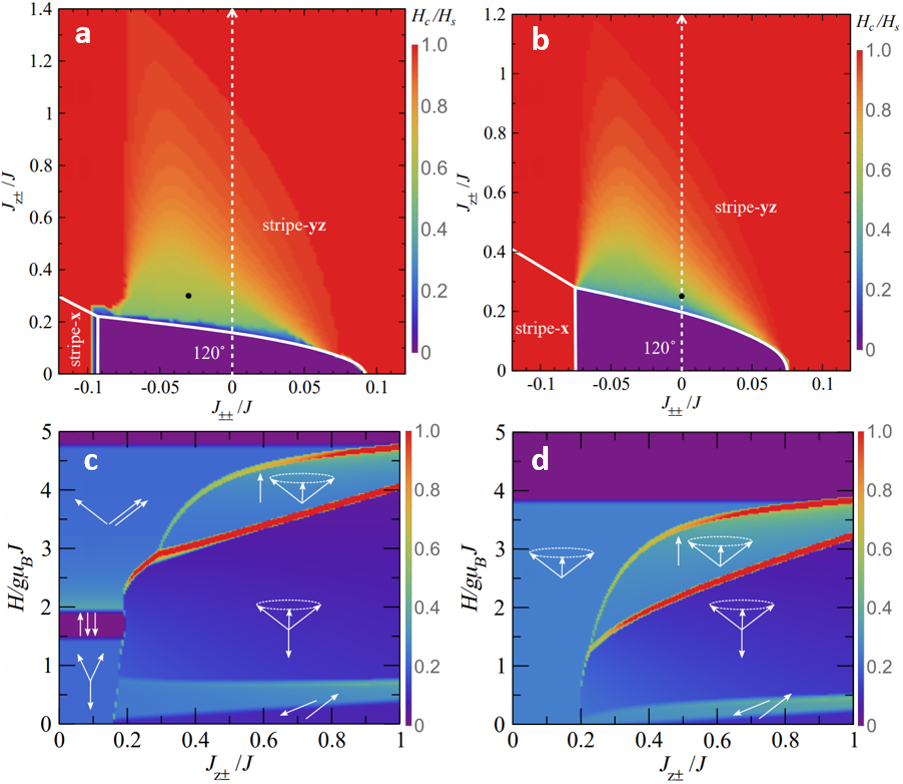}
\vskip -0.2cm
\caption{(a) Intensity plot of $H_c/H_s$ in $J_{z\pm}-J_{\pm\pm}$ axes for $\Delta =1.1$, $J_2=0.05J$. $H_c$ is the critical field of the transition from the four- to three-sublattice state, $H_s$ is the saturation field. Dot represents parameter set used in SpinW calculations. $H$ was scanned in steps of 0.1 and units of $g\muB J$, equivalent to steps of 0.021 in terms of $H_c/H_s$ for $H_s = 4.815$. (b) Same as in (a) for $\Delta =0.8$, where step size in $H_c/H_s$ was 0.026 corresponding to $H_s = 3.87$. (c) Intensity plot of the magnetic susceptibility, $\chi=dM/dH$, in the $J_{z\pm}$--$H$ plane
for $\Delta =1.1$, $J_2=0.05J$, and $J_{\pm\pm}=0$. Singularities correspond to phase transitions. (d) Same as in (c) for $\Delta =0.8$.}
\label{HcHsJzpJp}
\vskip -0.2cm
\end{figure*}

\subsection{Field-induced states}

To provide further insights into the effects of the field, we explore the new phases it can produce. In some cases, the field-induced transformation of the spin structures involves simple canting toward the field until a full saturation is reached at some $H_s$. However, in frustrated spin systems, or systems with anisotropic interactions, the field-evolution is more complicated. The case of the triangular-lattice Heisenberg antiferromagnet  is paradigmatic in this respect  \cite{starykh2015unusual}, showcasing a well-known and much-studied  sequence  of  transitions from ``Y'' to  ``up-up-down'' (UUD) plateau 
and ``V'' states in its field-evolution toward saturation. The $XXZ$ extension of the same model also includes non-coplanar ``umbrella'' and coplanar ``fan''  states  \cite{yamamoto2014quantum}. Next-nearest neighbor interactions and anisotropies also introduce a wider variety of four-sublattice field-induced states  \cite{ye2017half,seabra2011competition}.

However, the field-evolution of the phases  of the anisotropic-exchange model \eqref{HJpm}, which combines frustration from the bond-dependent terms  with that of the triangular-lattice geometry, remains largely unexplored. In this work, we offer some essential understanding and make significant steps of such an exploration. 

We use the same representative values of $J_2\!=\!0.05J$ and of the easy-axis and easy-plane $XXZ$ anisotropies  $\Delta\!=\!1.1$ and $\Delta\!=\!0.8$ as in Figs.~\ref{HcHsJzpJp} (a) and (b) to provide  an insight into the rich phase diagram of the model \eqref{HJpm}  in a field. In Figure ~\ref{HcHsJzpJp} (c) and (d),  we present intensity plots of the magnetic susceptibility $\chi\!=\!dM/dH$ in the $J_{z\pm}$--$H$ plane. They are obtained from the classical energy minimization of the four- and three-sublattice states in a field  for the two values of the $XXZ$ anisotropy and for $J_{\pm\pm}\!=\!0$.  Since  singularities in $\chi$ correspond to the phase transitions, these figures, in fact, constitute the 2D $J_{z\pm}$--$H$ phase  diagrams of the model (\ref{HJpm}) at fixed $\Delta$ and $J_2$ along the constant-$J_{\pm\pm}$ cut shown in Figs. \ref{HcHsJzpJp} (a) and (b).  Similar vertical cuts along $J_{z\pm}$ for the other values of  $J_{\pm\pm}$ and $J_2$ are also presented below to provide an understanding of the field-evolution of different states across the other dimensions of the phase diagram.

In case of the easy-axis $XXZ$ anisotropy in Figure~\ref{HcHsJzpJp} (c),  at the lower values of $J_{z\pm}$ one can observe the expected canonical sequence of the Y-UUD-V phase transitions of the three-sublattice states in the triangular-lattice $XXZ$ model  \cite{starykh2015unusual,yamamoto2014quantum}. Their corresponding spin structures are sketched in the figure.  

The field-induced behavior of the stripe states also includes multiple transitions. At the lowest field, the collinear stripe-\textbf{yz} spin configuration, in which spins are tilted off the basal plane of the lattice, is deformed into a non-coplanar four-sublattice state with all four spins on the elementary plaquette having different tilt angles. There is a broad crossover from this state to a structure with three spins forming an ``umbrella'' and the fourth strictly antiparallel to the magnetic field, see the sketches of the spin order in Figure~\ref{HcHsJzpJp} (c). This latter state is stable in a wide field region. As the field increases further, at  not too small $J_{z\pm}$ there is a spin-flop-like transition to a similar state, umbrella with a fourth spin parallel to the field. For the yet larger values of $J_{z\pm}$,  transition to the saturation occurs directly from this ``umbrella+up" state 
via a first-order transition.

The main feature in both Figs. \ref{HcHsJzpJp} (c) and (d), which is also our key finding, is that the region of stability of the three-sublattice states, related to the experimentally observed intensity at the $K$ point, \textit{expands} at the larger values of the magnetic field. Therefore, there is a region of the model parameters where an evolution from the four-sublattice to the saturated state \emph{necessarily} proceeds via a high-field three-sublattice state. 

For the easy-axis  case of Figure~\ref{HcHsJzpJp} (c), this high-field state is a coplanar ``V'' state. For the easy-plane $XXZ$ anisotropy case of Figure~\ref{HcHsJzpJp} (d), the four-sublattice phases and all the discussed trends are the same, while the three-sublattice region, classically, is a single noncoplanar ``umbrella'' state, which is a canted $120^\circ$ structure. We note that in the quantum case and for not too small $XXZ$ anisotropy $\Delta$, this umbrella state is replaced by the same sequence  of Y-UUD-V phases as in Figure~\ref{HcHsJzpJp} (c)   \cite{yamamoto2014quantum}. In contrast with the large-$J_{z\pm}$ first-order transition  from the four-sublattice state to saturation, the transition from both V and umbrella states to saturation is second-order. We note that the discussed  transitions agree with the prior classical Monte-Carlo simulations  \cite{steinhardt2020fieldinduced} conducted in the context of the parameter search for \ymgo for a narrower range of parameters. 

\begin{figure*}
\includegraphics[width=\linewidth]{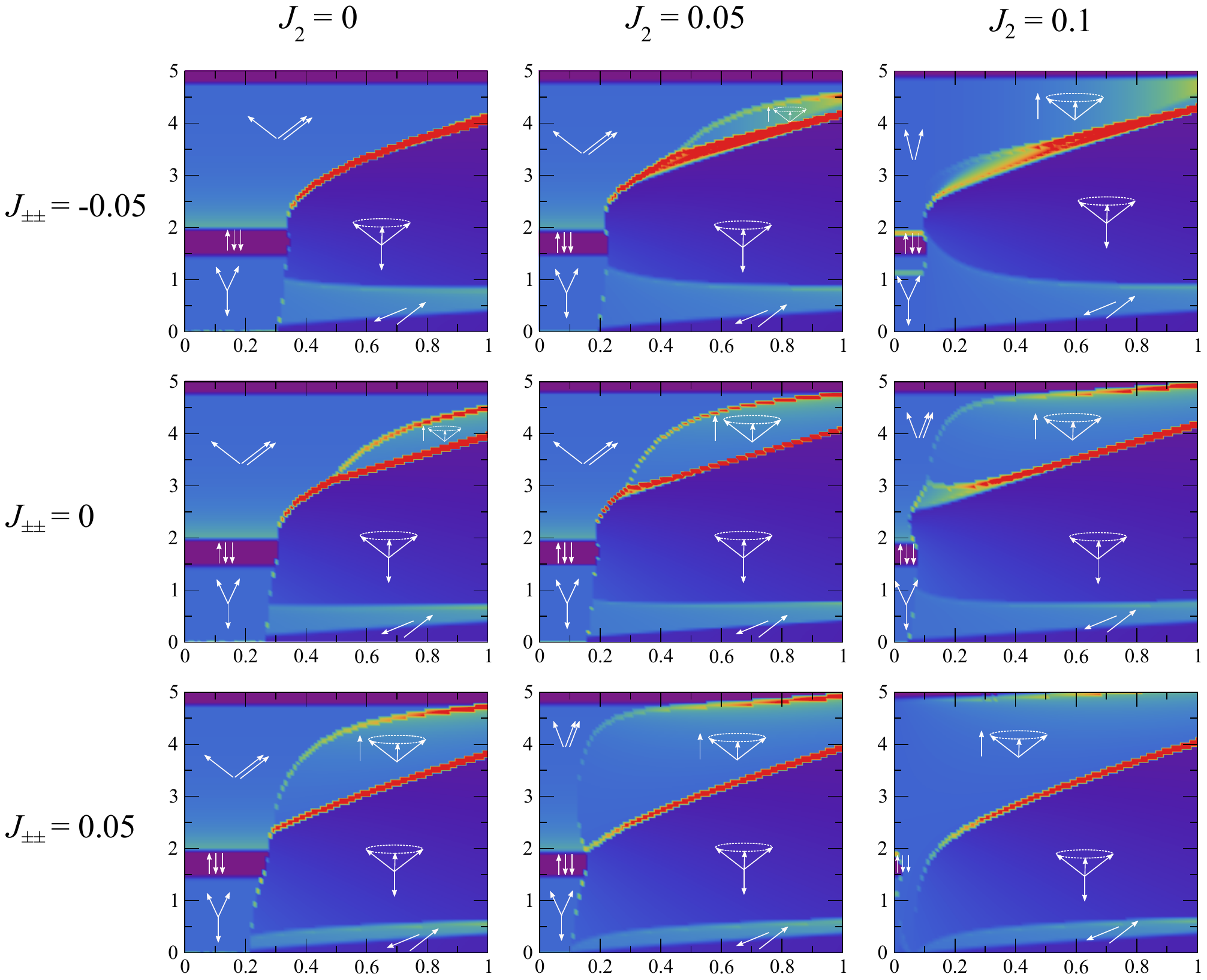}
\vskip -0.2cm
\caption{Intensity plots of the magnetic susceptibility for the easy-axis case, $\Delta =1.1$ 
and for various $J_2$ and $J_{\pm\pm}$. The axes and color-scale are the same as in Figure~\ref{HcHsJzpJp} (c), which is
also the central panel here.}
\label{fig_d11_j2_jpp}
\end{figure*}   

\subsection{Further evolution of the  phases}

\begin{figure*}
\includegraphics[width=\linewidth]{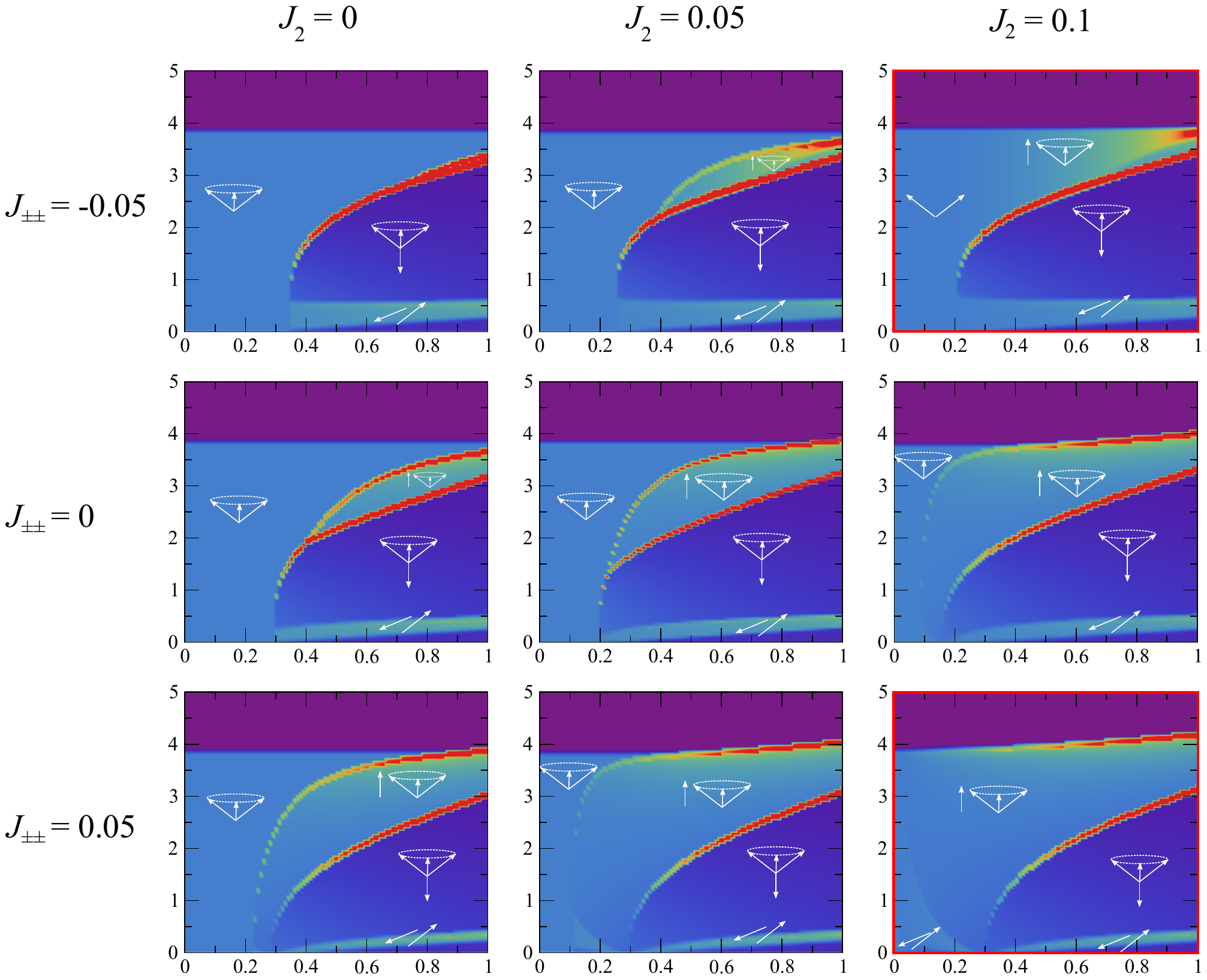}
\caption{Same as in Figure~\ref{fig_d11_j2_jpp} for  $\Delta =0.8$.
Phase diagrams without three-sublattice states are marked with the red frame.}
\label{fig_d08_j2_jpp}
\end{figure*}

Here we present further details on the field-evolution of  various phases of the model (\ref{HJpm}) for  different choices of $J_2$ and $J_{\pm\pm}$. Our Figs.~\ref{fig_d11_j2_jpp} and \ref{fig_d08_j2_jpp} show the intensity plots of the magnetic susceptibility in the $J_{z\pm}$--$H$ plane for the same choices of the $XXZ$ anisotropy  $\Delta=1.1$ and $\Delta=0.8$ as in Figs.~\ref{HcHsJzpJp} (c) and (d), respectively. Each row of the graphs corresponds to a different constant-$J_{\pm\pm}$ cut of Figs.~\ref{HcHsJzpJp} (a) and (b) and each column corresponds to a different  constant-$J_2$ cut of the 3D phase diagram in Figure~\ref{fig_3d}.

While the variations of the $J_{\pm\pm}$ term simply give an elaborate dissection of what happens underneath the projected view of the intensity plots of Figure~\ref{HcHsJzpJp} (a) and (b) for $J_{\pm\pm}=\pm 0.05J$ cuts, it is clear that the next-nearest-neighbor interaction $J_2$ works strongly against the field-induced three-sublattice states. This is in accord with  Figure~\ref{fig_3d}, which shows that at  $J_2\approx0.125J$ the $120^{\degree}$ state is completely eliminated from the zero-field phase diagram. The sets of parameters where no three-sublattice state is observed at any field are highlighted with a red frame in Figure~\ref{fig_d08_j2_jpp}. There are other changes that include shrinking or expanding of the phases already described and the appearance of a new canted stripe state for large $J_2$ and negative $J_{\pm\pm}$ that occur at  $J_{z\pm}\lesssim 0.2J$ and larger fields for $\Delta\!=\!1.1$ and for all fields  for $\Delta\!=\!0.8$, see upper right panels of Figs.~\ref{fig_d11_j2_jpp} and \ref{fig_d08_j2_jpp}.

The main takeaway from  Figs.~\ref{fig_d11_j2_jpp} and \ref{fig_d08_j2_jpp} is that the region of the four- to three-sublattice transition is limited by the extent of $J_{\pm\pm}$ already shown in  Figs.~\ref{HcHsJzpJp} (a) and (b) and more strongly by the next-nearest-neighbor interaction $J_2$. Therefore, the experimental observations that the critical field of such a transition $H_c\!\alt\!H_s$ in both YMGO and YZGO put strong bounds on the anisotropic-exchange parameters. This analysis also limits next-nearest neighbor interactions to $J_2\!\alt\!0.1J$ as only four-sublattice 
states survive for larger $J_2$.

\subsection{$S({\bf Q},\omega)$ for select parameters}
In order to relate the above theoretical analysis to our experimental results for \yzgo and \ymgo, we used SpinW \cite{toth2015linear} to calculate linear spin-wave theory (LSWT) $S({\bf Q},\omega)$ for representative sets of parameters from the identified regions of the phase diagram that are shown by dots in Figs.~\ref{HcHsJzpJp} (a) and (b). Since the phase diagrams are for the parameters in units of the overall scale $J$, the latter is set to match  LSWT dispersion in high fields.  Comparison of the data to calculations from SpinW is inherently challenging owing to the implicit assumption in the SpinW calculations of the absence of disorder (which is necessary to reproduce the broadening in energy and $Q$ observed in experiment).  Nonetheless, for the parameters indicated in Figure \ref{HcHsJzpJp} (a) we find very good qualitative agreement between calculations and data (see Figure \ref{CNCS}) in the absence of disorder.  The SpinW optimization algorithm optmagsteep was used with iterative manual adjustment of the spin state until the stable spin state (free of imaginary modes) was found\footnote{We should note that the ``V'' state spectrum is unstable for 2.5 T near the $\Gamma$ point for the chosen set of parameters. The instability is weak and may signify a transition to a more complicated multi-$\mathbf{Q}$ state. While the manual tweaking allows us to find a quasi-stable state with the gapped spectrum, the stable ``V'' state spectrum is gapless as any three-sublattice state of \eqref{HJpm} with broken continuous symmetry, see also Ref.\cite{maksimov2019anisotropic} for the $120^\circ$ state spectrum.}.  

In particular, comparing the diffuse scattering from Figure \ref{CNCS} (a) 0 T and the resolution-convoluted dispersion calculated from SpinW in Figure \ref{CNCS} (f), we note the enhanced intensity in the vicinity of M and K clearly present in both.  Furthermore, the greater intensity residing in a range of energy from near the elastic line to about 0.2 meV in the calculation is clearly present, albeit undoubtedly broadened by disorder, in the experimental data.  As discussed in previous works \cite{paddison2017continuous,ma2018spin,li2017crystalline}, disorder ``smears'' the intensity in $Q$ and $E$, which would also account for the intensity visible at higher energies in the INS data.  

In Figure \ref{CNCS} (g) we show the calculated dispersion using SpinW at 2.5 T, where the spins reside in the `V' state.  The dispersion has again been convoluted with the resolution of the instrument for best comparison to the experimental data.  Note that the intensity at the $\Gamma$ point (though suppressed as discussed below) is at approximately 0.5 meV, as expected from comparison to the 2 and 3 T data.  Furthermore, the modes above the Brillouin zone edge (which we emphasize follow from the superposition of 3 domains in the clean limit) offer a hint as to the the origins of the the broad continuum apparent in the data.  As the field is raised, the contributions from modes above the zone edge in the SpinW calculation are reduced (compare the intensity from modes above M and K at 0 and 2 T).  When the system is fully saturated, and the `V'-state gives way to the field-induced ferromagnetic state, the modes above the zone edge are completely absent (see the curve calculated from the analytic expression following from linear spin-wave theory superimposed on the 8 T INS data in Figure \ref{CNCS} (h)).  This evolution of intensity is qualitatively consistent with that observed in the data, where broadening due to disorder smears out the modes.  That is, the modes residing above the zone edge contribute to the continuum at low to intermediate field, but are reduced in intensity as the system enters the (subject-to-disorder) `V'-state, and are replaced by the broadened, otherwise singular mode when the system is fully polarized.

We also point out that an apparent visibility of the magnon modes in LSWT $S({\bf Q},\omega)$ in 
the vicinity of the $\Gamma$ point at low fields, Figure~\ref{CNCS}(f),(g), and lack thereof in the experimental data, Figure~\ref{CNCS}(a-c), can be explained by a significant interaction of the single-magnon branch with the two-magnon continuum. While the quantitative calculation of such effects in the anisotropic-exchange models is quite involved \cite{Kopietz2020, winter2017breakdown} and is beyond the scope of the present work, we, nevertheless, provide an intuitive insight into it by showing the bottom of the two-magnon continuum for three magnetic domains in Figure~\ref{CNCS}(f) and (g). As shown by the dashed lines in Figure~\ref{CNCS} for both 0 T and 2.5 T, the bottom of the continuum has lower energy than one magnon modes at the large portion of the Brillouin zone, including the $\Gamma$ point. For $S=1/2$ systems, such an overlap can lead to strong decays and near-complete disappearance of the well-defined 
magnon excitations in the corresponding range of momenta together with the other phenomena such as strong renormalization of  the single-magnon branches \cite{zhitomirsky2013colloquium,chernyshev2009spin}. These effects require significant coupling between one- and two-magnon sectors - an inevitable consequence of the anisotropic exchange terms in \eqref{HJpm}, which are allowed by the presence of strong spin-orbit coupling in the rare-earth-based and some transition-metal compounds\cite{maksimov2020rethinking,winter2017models,Kopietz2020}.

We have also calculated the 0 T $S(\vec{Q},\omega)$ for the parameters identified as appropriate for the easy-plane character of \ymgo, identified by a dot in Figure~\ref{HcHsJzpJp} (b), (see Supplementary Figure 10), which is in excellent qualitative agreement with Figure 2 (a) of  \cite{paddison2017continuous}.  We further calculated the $S(\vec{Q},\omega)$ for the field-induced polarized state, where the overall scaling of the model (given by $J$) was chosen for best agreement.  Within the uncertainty of the broad scattering observed in experiment, this calculation also yields a very good qualitative agreement (Supplementary Figure 11).  
\section{Discussion}
The hunt for the QSL state has yielded numerous studies of interesting new compounds and underlying physical phenomena, even though the ultimate goal may remain elusive.  The search for a QSL state in the context of \yzgo and \ymgo materials seems \emph{likely} nearing its end, although one cannot claim yet with absolute certainty that the issue is completely settled \cite{wu2020exact}. In the present work, we have established a potentially close description of \yzgo and \ymgo in a hypothetical \emph{clean} limit, allowing for the implications of the disordered, real-world materials to follow.  While possibly denied the coveted QSL state, a deeper understanding of the rich and diverse phase diagram of the triangular-lattice antiferromagnets is still a highly satisfying reward that will likely serve as a road map for future studies, in particular for designing new materials or advancing these studies to explore this rich phase diagram. 

The present work seeks to advance our understanding of the phase diagram describing \ymgo, \yzgo, and many related systems.  We note that our specific conclusions about these materials are a part of a greater effort to narrow down  possible magnetic exchange parameters in these systems  \cite{zhang2018hierarchy,bachus2020field,li2020reinvestigation,steinhardt2020fieldinduced}, and provide a guideline on how to address the issue of chemical disorder in real-world materials.  In a very recent work \cite{steinhardt2020fieldinduced} by some of the same authors,  the phase crossovers similar to those shown in Figure \ref{TDO_SQUID} and Figure \ref{CORELLI} were used to constrain parameters of \ymgo, with the focus on reproducing them in the disorder-free limit, but \emph{without} a broader map of the phase diagram beyond the observations offered by an optimization algorithm. Indeed, the methods used in that prior work suggest a general approach to find exchange parameters for disordered systems.  By contrast, the goal of \emph{this} work is to provide an expanded context of the phase diagram in which the observed crossovers \emph{can occur in principle}, offering a  common description for the phenomena observed in both systems. As such, the present work will serve as an invaluable map in the search of QSL  and other intriguing states and phenomena in frustrated triangular-lattice compounds, and will importantly benefit the materials design efforts in this field. 

With this last point in mind,  recent experiments in the ytterbium-based chalcogenides also show promising QSL features and numerous phase transitions induced by an applied magnetic field \cite{liu2018rare,ding2019gapless,ranjith2019anisotropic,ranjith2019field,baenitz2018planar,xing2019field}. The precise nature of these   transitions is not yet known and needs to be analyzed in a manner and within the framework suggested in the present study. These recent studies also highlight potential further insights that can bring deeper understanding of \yzgo and \ymgo by experiments  with the field along different directions. 

Altogether, we have demonstrated the power of the experimental and theoretical insights in identifying relevant parameter spaces of \ymgo and \yzgo and, potentially, other related materials.  We have investigated their field-induced phases  and have shown that despite their disorder-induced pseudo-SL ground states, we can significantly narrow the allowed regions of their phase diagram that are compatible with the phenomenologies of these materials. Similar consideration can be applied to the other materials such as chalcogenides. More experimental and theoretical investigations, such as targeted materials design, neutron scattering for the in-plane field directions augmented by analytical and numerical studies for this setting, and utilizing alternative tuning parameters such as external pressure, all could be useful for exploring the diverse phase diagram of this group of compounds and shedding further light on their rich physics.

\section{Methods}
\subsection{Synthesis}
Samples were synthesized from finely mixed Yb$_2$O$_3$ (99.9$\%$), ZnO(99.9$\%$), and Ga$_2$O$_3$ (99.999$\%$) powders at 1350 $\degree$C.  High quality single-crystals (such as pictured in Supplementary Figure 2) were grown using the optical floating zone technique.  A typical growth was conducted in 1 MPa O$_2$ atmosphere with a speed ranging from 4 to 10 mm/hour.  Single crystal quality was confirmed via laue x-ray diffraction (Supplementary Figure 3) while powder xray diffraction (PXRD) was used to confirm the correct phase at every step of synthesis (see Supplementary Figure 1).  

\subsection{High Resolution Magnetization Measurements}
High-resolution measurements of magnetization were achieved with the complimentary tunnel diode oscillator (TDO) technique  \cite{steinhardt2020fieldinduced}. In a TDO measurement, a tunnel diode is biased to operate in the “negative resistance” region of the IV-curve. This provides power that maintains the resonance of a LC-circuit at a frequency range between 10 and 50 MHz. A nearly single-crystal sample with dimensions of ~2 mm in length and ~1 mm in diameter was wound inside a detection coil, with the $\textbf{c}$ axis of the sample aligned with the coil axis. The sample and coil constitute the inductor of the LC circuit. With the application of field, the sample magnetization change induces a change in the inductance, and thus shifts the resonance frequency. This technique enables highly sensitive detection of changes of magnetic moments ~ $10^{-15}$ Am$^2$.  \cite{van1975tunnel}. 

Magnetization was also directly measured via an in-house Cryogenic S700X SQUID magnetometer in temperatures down to 300 mK using a 3He probe.  A 1.81 mg sample was mounted on a silver straw with vacuum grease in $\textbf{H} \parallel \textbf{c}$ and $\textbf{H}\perp\textbf{c}$ orientations.   

ac-susceptibility measurements were carried out on a long, approximately rectangular sample with field perpendicular to the sample $c$ axis in a dilution refrigerator with a base temperature of 20 mK.  

\subsection{Neutron Scattering}
\subsubsection{Diffuse magnetic scattering at CORELLI}
Diffuse neutron scattering data were collected at the CORELLI spectrometer at Spallation Neutron Source, Oak Ridge National Laboratory \cite{rosenkranz2008corelli}. This instrument is a quasi-Laue TOF instrument equipped with a 2D detector, with a -20$\degree$ to +150$\degree$ in-plane coverage. The incident neutron energy was between 10 meV and 200 meV. A superconducting magnet was used to provide a vertical magnetic field up to 5 T, which constrained the out-of-plane coverage to $\pm$ 8$\degree$. A ~0.8 g single crystal was mounted on a Cu plate in a dilution refrigerator. The sample was aligned with the $(\textbf{h}, \textbf{k}, 0)$ plane horizontal and the magnetic field along the [0,0,$\textbf{l}$] direction. Neutron-absorbing Cd was used to shield the sample holder to reduce the background scattering. Experiments were conducted with applied fields at the base temperature of 130 mK by rotating the crystal through 180$\degree$ in 3$\degree$ steps, and then at 20 K in the same fields for background subtraction. The data were reduced using Mantid for the Lorentz and spectrum corrections \cite{michels2016expanding}.

To account for the temperature factor in our background subtraction in total-scattering mode, we compared the ratio of integrated intensities of a rectangular volume of reciprocal space at 5 T for both temperatures.  The region was bounded by $-0.45 < \textbf{h} < -0.35$, $0.5 < \textbf{k} < 0.6$, and $-2 < \textbf{l} < 2$.  We used this region, away from the zone edge, and our 5 T data to ensure the comparison was unaffected by diffuse magnetic scattering.  This integration approximates the ratio of Bose population factors - our scaling factor was $1.01\pm0.005$. To improve statistics, we used symmetry operations.  All analysis and visualization were performed using Mantid and Python.

\subsubsection{Inelastic neutron scattering at the CNCS, DCS, and SPINS}
We conducted inelastic neutron scattering experiments at the Cold Neutron Chopper Spectrometer\cite{ehlers2011cncs} (CNCS) at Oak Ridge National Laboratory, and the Disk-Chopper Spectrometer\cite{COPLEY2003477} (DCS) and Spin Polarized Inelastic Neutron Spectrometer (SPINS) at the National Institute of Standards and Technology.  The same ~1.4 g single crystal sample was aligned to use the $[h,k,0]$ scattering plane and with the field parallel to the sample $\vec{c}$ axis (along the [001] direction).  All measurements were carried out in dilution refrigerators, with sample mounted on copper sample mounts such as pictured in Supplementary Figure 2. 

The base temperatures for CNCS, DCS, and SPINS measurements were 50 mK, 70 mK, and 60 mK, respectively.  For CNCS and DCS, the sample was rotated through approximately 180 degrees.  The incident energy used at CNCS, DCS, and SPINS was 3.9 meV, 3.55 meV, and 3.7 meV, respectively.

Analysis of CNCS data was carried out in large part using the HORACE software package\cite{Ewings2016132}.  Analysis of DCS data was carried out in large part using the DAVE software package\cite{azuah2009dave}.  

\subsubsection{Polarized neutron scattering at BT7}
We measured using polarized neutrons at the BT7\cite{lynn2012bt7} beamline at the National Instistute of Standards and Technology.  We measured with a fixed final energy of 14.7 meV at about 0.5 meV such that the FWHM of the collimated beam was 1.08 meV and encompassed the energy range of the continuum as pictured for low fields in Figure \ref{CNCS}.  Samples were again aligned to use the [h,k,0] scattering plane, and measurements were conducted using a vertical guide field (parallel to the sample c-axis).  A 3He cryostat was used.  0 T measurements were performed in the absence of a magnet, and then a 7 T magnet was added to perform measurements at 2 T.  Flipping ratios ranged from 19 to 33 for the 0 T measurements and from ~17 to ~28 for the 2 T measurements.  Polarization correction was performed using \emph{pbcor} software.

\label{sec:methods}

\subsubsection{Theory}

For the LSWT $S({\bf Q},\omega)$  we used  SpinW calculations \cite{toth2015linear}. The global phase diagram was obtained using classical energies for the single-${\bf Q}$ states, see Refs.~\citep{zhu2018topography,maksimov2019anisotropic}, and for the field-induced phases, classical energy
minimization of the three- and four-sublattice structures was used. 

\section{Data availability}
All relevant data are available from the authors upon reasonable request. This manuscript has been authored by UT-Battelle, LLC under Contract No. DE-AC05-00OR22725 with the U.S. Department of Energy. The United States Government retains and the publisher, by accepting the article for publication, acknowledges that the United States Government retains a non-exclusive, paid-up, irrevocable, world-wide license to publish or reproduce the published form of this manuscript, or allow others to do so, for United States Government purposes. The Department of Energy will provide public access to these results of federally sponsored research in accordance with the DOE Public Access Plan(http://energy.gov/downloads/doe-public-access-plan).

\begin{acknowledgments}
The work of A.~L.~C. was supported by the U.S. Department of Energy, Office of Science, Basic Energy Sciences under Awards No. DE-FG02-04ER46174 and DE-SC0021221. P.~A.~M. acknowledges support from JINR Grant for young scientists 20-302-03. A.~L.~C. would like to thank  Kavli Institute for Theoretical Physics (KITP) where this work was advanced. KITP is supported  by the National Science Foundation under  Grant No. NSF PHY-1748958.  A portion of this work was performed at the National High Magnetic Field Laboratory, which is supported by the National Science Foundation Cooperative Agreement No. DMR1157490 and DMR-1644779, the State of Florida and the U.S. Department of Energy.  A portion of this research used resources at the Spallation Neutron Source, a DOE Office of Science User Facility operated by the Oak Ridge National Laboratory. We acknowledge the support of the National Institute of Standards and Technology, U.S. Department of Commerce, in providing the neutron research facilities used in this work.  The identification of any commercial product or trade name does not imply endorsement or recommendation by the National Institute of Standards and Technology.
\end{acknowledgments}

\section{Author Contributions}
Research conceived by S.H.; Samples synthesized by W.S., C.M. and S.H.; Magnetic measurements performed and analyzed by Z.S., W.S., D.G., and S.H.; Neutron scattering measurements performed and analyzed by W.S., S.D., N.P.B., A.P., Y.L., G.X., Y.Z., J.W.L, and S.H.; Theoretical calculations performed by P.A.M and A.L.C.; Manuscript written by W.S., P.A.M, A.L.C., and S.H.; Project supervised by A.L.C and S.H.; All authors commented on the manuscript.

\section{Competing Interests}
The Authors declare no Competing Financial or Non-Financial Interests.

\bibliographystyle{apsrev4-1}
\bibliography{1_main.bbl}
\clearpage
\setcounter{figure}{0}

\newpage
\onecolumngrid
\begin{center}
{\large\bf Phase Diagram of \yzgo in Applied Magnetic Field \\ \emph{Supplementary Materials}}\\ 
\vskip0.35cm
William Steinhardt,$^1$ P. A. Maksimov,$^2$ Sachith Dissanayake,$^3$ Nicholas P. Butch,$^3$ David Graf,$^4$, Andrey Podlesnyak,$^5$ Yaohua Liu,$^5$, Yang Zhao,$^{3,6}$, Guangyong Xu,$^3$ Jeffery W. Lynn,$^3$, Casey Marjerrison,$^1$ A. L. Chernyshev$^7$, and Sara Haravifard$^{1,8}$\\
\vskip0.15cm
{\it \small $^1$Department of Physics, Duke University, Durham, North Carolina, 27008, USA}\\
{\it \small $^2$Bogolyubov Laboratory of Theoretical Physics, Joint Institute for Nuclear Research, Dubna, Moscow region 141980, Russia}\\
{\it \small $^3$NIST Center for Neutron Research, National Institute for Standards and Technology, Gaithersburg, Maryland, 20899, USA}\\
{\it \small $^4$National High Magnetic Field Laboratory and Department of Physics, Florida State University, Tallahassee, Florida, 32310, USA}\\
{\it \small $^5$Neutron Scattering Division, Oak Ridge National Laboratory, Oak Ridge, Tennessee, 37831, USA}\\
{\it \small $^6$Department of Materials Science and Engineering, University of Maryland, College Park, Maryland, 20742, USA}\\
{\it \small $^7$Department of Physics and Astronomy, University of California, Irvine, California, 92697, USA}\\
{\it \small $^8$Department of Mechanical Engineering and Materials Science, Duke University, Durham, North Carolina 27708, USA}\\
\vskip 0.1cm \
\end{center}
\twocolumngrid

\begin{figure*}
\includegraphics[width=12cm]{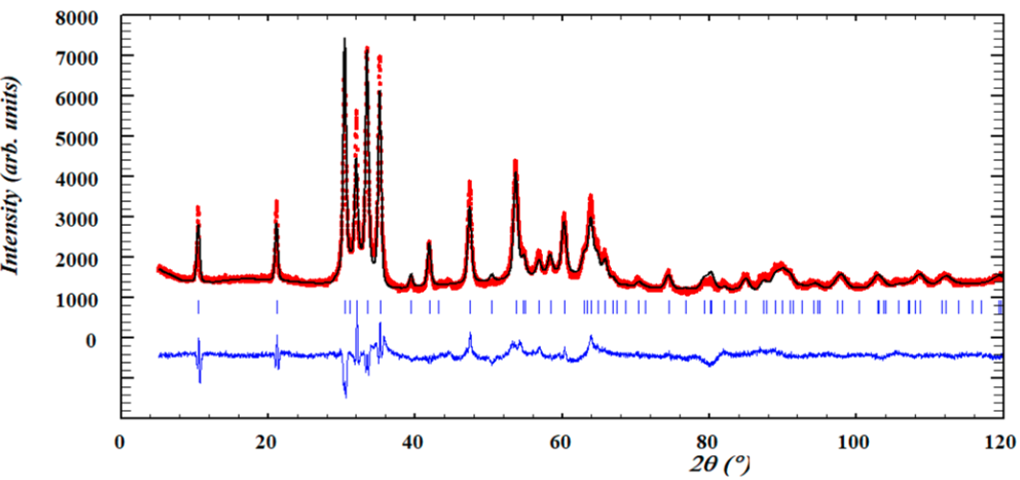}
\vskip -0.2cm
\caption{Powder x-ray diffraction confirms the correct phase of the sample.} 
\label{PXRD}
\end{figure*}

\begin{figure*}
\includegraphics[width=8cm]{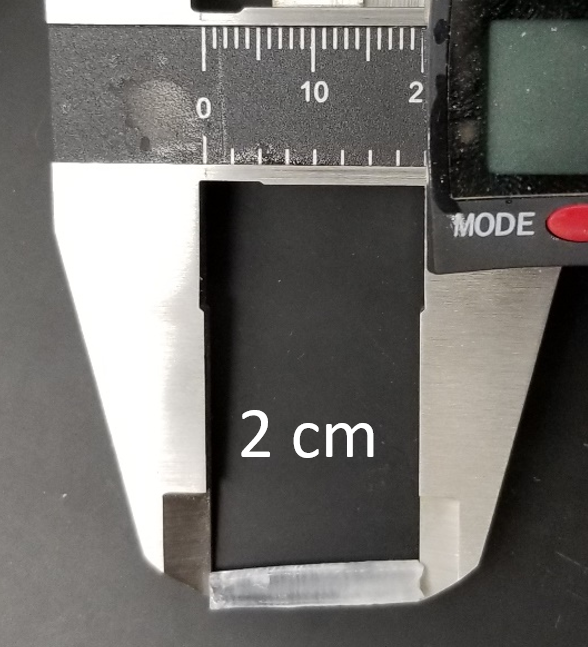}
\includegraphics{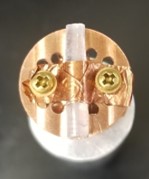}
\vskip -0.2cm
\caption{The optical floating zone technique was used to produce large, high-quality single crystal sample.  Left, the crystal pictured with caliper for scale.  Right, crystal mounted on a copper plate for applied field neutron scattering experiments.} 
\label{crystal}
\end{figure*}

\begin{figure*}
\includegraphics[]{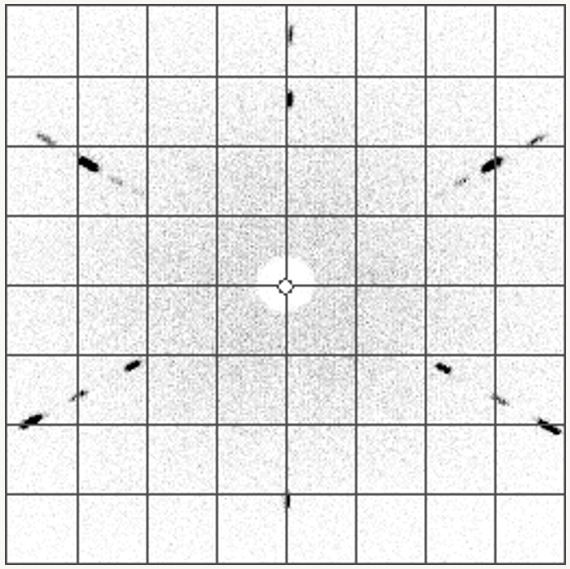}
\vskip -0.2cm
\caption{Laue x-ray diffraction measured with backscattering from the basal planes confirm the high quality and single-crystalline nature of the sample.} 
\label{laue}
\end{figure*}

\begin{figure*}
\includegraphics[width=16cm]{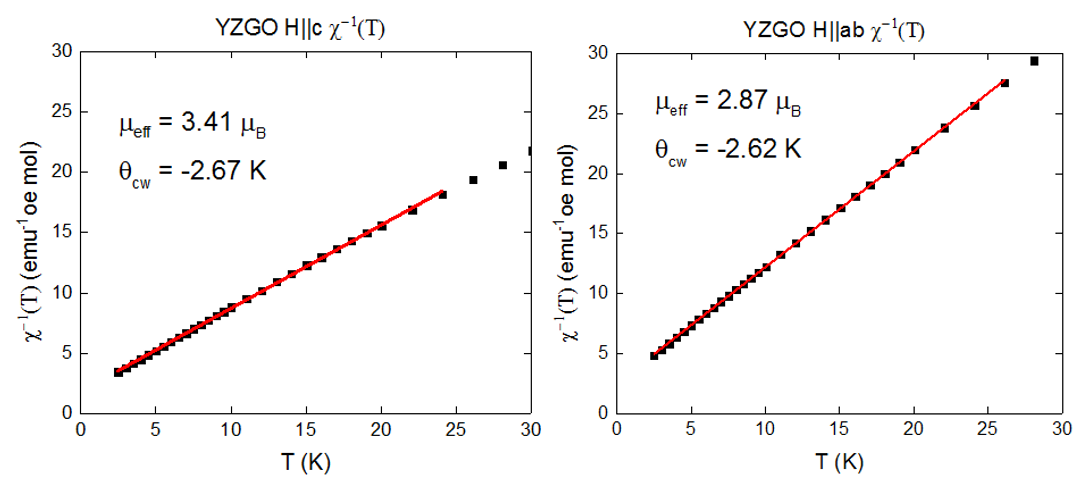}
\vskip -0.2cm
\caption{Inverse susceptibility is plotted versus temperature, indicating the Curie-Weiss temperatures shown for two orientations of the crystals with respect to field. Uncertainties represent one standard deviation.  Note: 1 emu/(mol$\times$Oe) = 4$\times$ 10$^{-6}$ m$^3$/mol.} 
\label{inverse}
\end{figure*}

\begin{figure*}
\includegraphics[width=6cm]{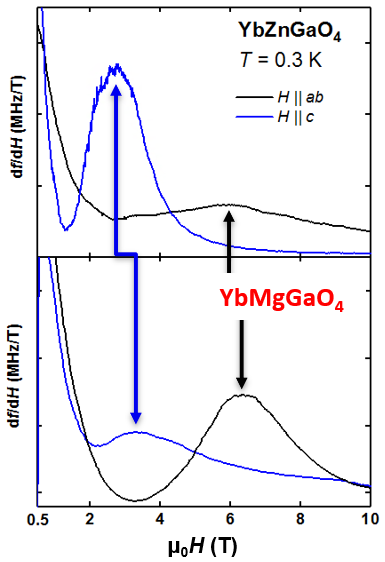}
\vskip -0.2cm
\caption{Comparison of the first derivative of the change in frequency measured with TDO with respect to field for YbZnGaO$_4$ (above) and YbMgGaO$_4$ (below).  The onset of the anomaly detected in both measurements appears lower for YbZnGaO$_4$ for both orientations, where arrows have been added to guide the eye.} 
\label{compare_tdo}
\end{figure*}

\begin{figure*}
\includegraphics[width=16cm]{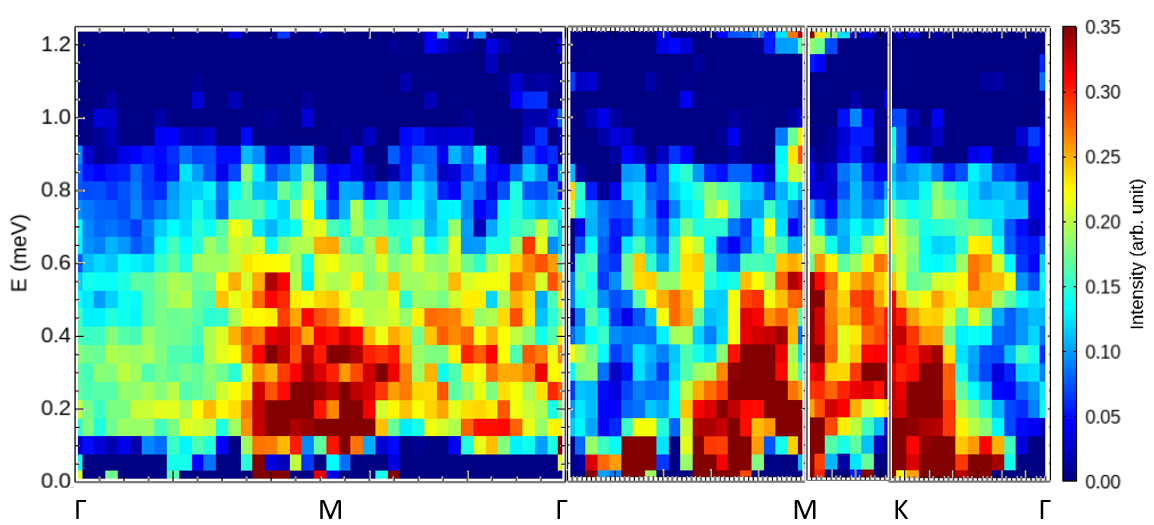}
\vskip -0.2cm
\caption{INS scattering data collected along the same path as shown in Figure 3 (a) in the main text at 70 mK at the disk chopper spectrometer at NIST at 0 T, with 8 T data subtracted as background.  Features are in very good agreement with data collected at CNCS.} 
\label{DCS_0T}
\end{figure*}

\begin{figure*}
\includegraphics[width = 18cm]{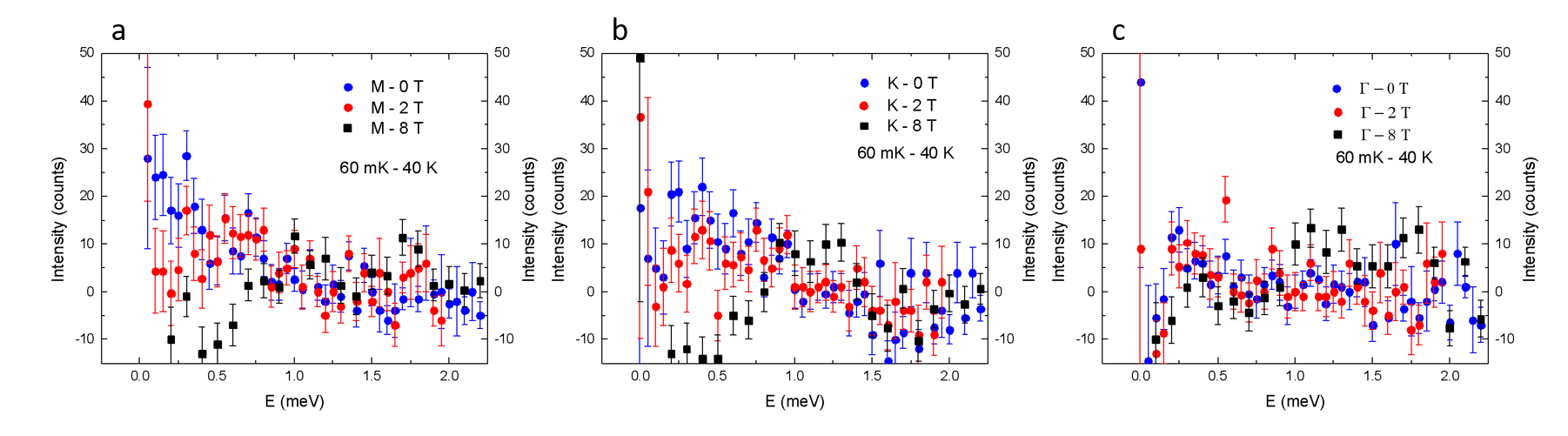}
\vskip -0.2cm
\caption{Inelastic neutron scattering data collected at 60 mK at the instrument SPINS at NIST.  These constant Q cuts for the high-symmetry points corroborate observations at DCS and CNCS. Uncertainties represent one standard deviation.} 
\label{spinsfield}
\end{figure*}

\begin{figure*}
\includegraphics[width=12cm]{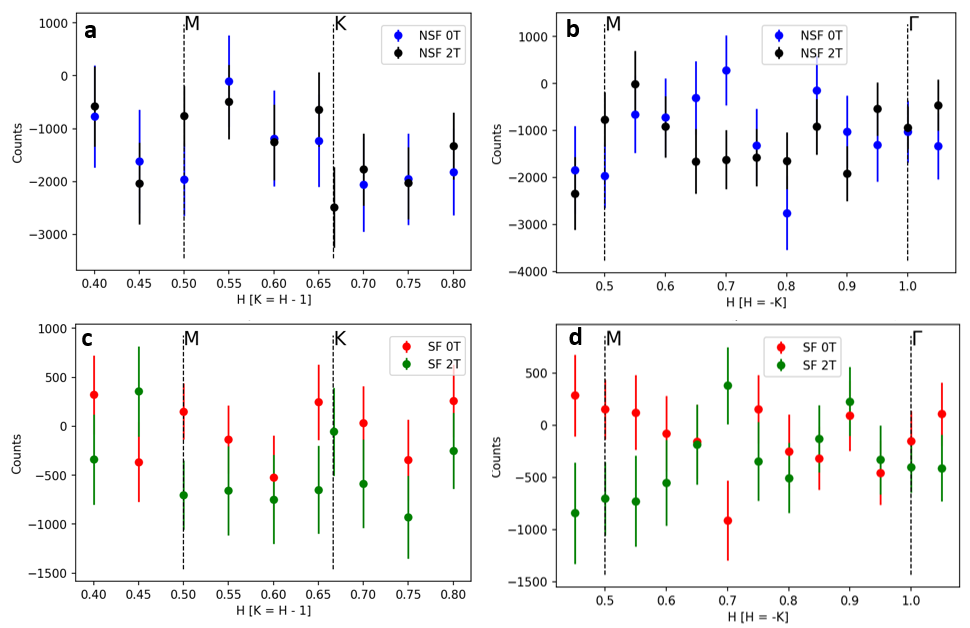}
\vskip -0.2cm
\caption{Polarized neutron scattering in vertical polarized mode.  0.3 K with 40 K background subtracted. a) Non-spin-flip 0 T and 2 T data collected along the Brillouin zone shows generally greater intensity for 2 T (corresponding to the enhanced projection out-of-plane for higher fields)  b) As in a) but for the cut extending from M to $\Gamma$.  At low fields, most scattering is along the zone edge.  c) As in a) but showing the spin-flip channel for 0 T and 2 T, showing greater intensity for in-plane components, consistent with a). d) As in b) but showing the spin-flip channel for 0 and 2 T.  See methods for details. Uncertainties represent one standard deviation.} 
\label{polarized}
\end{figure*}

\begin{figure*}
\includegraphics[width=12cm]{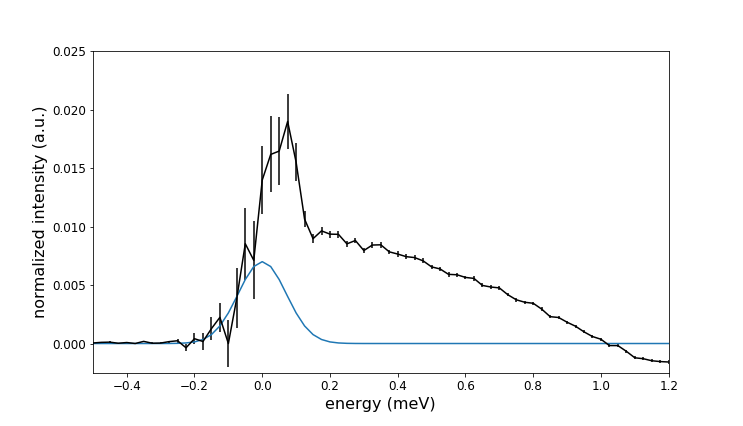}
\vskip -0.2cm
\caption{Intensity versus energy at the M point using 0 T - 8 T subtraction (both base temperature).  Blue curve shows estimated scattering from the elastic line based on the 0.141 meV instrument resolution at the elastic line, which is approximately 16\% of the total scattering.  Over-subtraction at ~1 meV is due to the single-magnon dispersion present in the 8 T data. Uncertainties represent one standard deviation.} 
\label{frozen}
\end{figure*}
\begin{figure*}
\includegraphics[width=12cm]{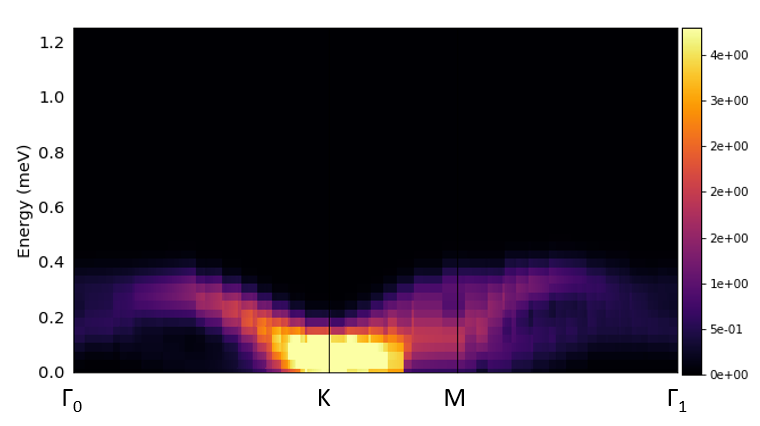}
\vskip -0.2cm
\caption{S(Q,$\omega$) calculated using SpinW \cite{toth2015linear} for parameters identified in the main text for YbMgGaO$_4$ for no applied field.  Calculation is for the material in the disorder-free limit. Note good qualitative agreement with Figure 2a of \cite{paddison2017continuous}.  Parameters used for SpinW calculations in are (in meV) $J=0.218$, $J_{\pm\pm} = 0$, $J_{z\pm} = 0.0545$, $J_{zz}=0.1744$, $J^2=0.01095$, $J_{zz}^2=0.0087$, $g_{\parallel}=3.82$, such that  $J_{\pm\pm}/J = 0$, and $J_{z\pm}/J=0.25$. }  
\label{ymgospinW0T}
\end{figure*}

\begin{figure*}
\includegraphics[width=12cm]{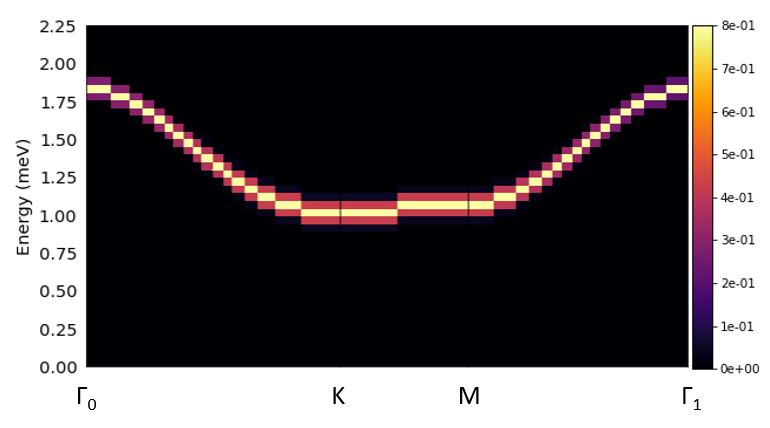}
\vskip -0.2cm
\caption{S(Q,$\omega$) calculated using SpinW \cite{toth2015linear} for parameters listed in previous caption and identified in the main text for YbMgGaO$_4$ for the field-induced polarized state at 8 T.  Calculation is for the material in the disorder-free limit. Note good qualitative agreement with Figure 3a of \cite{zhang2018hierarchy}.  Parameters used for calculation are identical to those listed in SM Fig. 10 caption.} 
\label{ymgospinW8T}
\end{figure*}

\bibliographystyle{apsrev4-1}
\clearpage

\end{document}